\pdfoutput=1

\documentclass[12pt]{article}

\usepackage{amsfonts}\usepackage{amsmath}\usepackage{amssymb}\usepackage{bigints}
\usepackage{booktabs}\usepackage[nosort]{cite}\usepackage{color}\usepackage{dsfont}
\usepackage{float}\usepackage{framed}\usepackage{graphicx}\usepackage{indentfirst}
\usepackage{mathrsfs}\usepackage{multirow}\usepackage{pdflscape}\usepackage{setspace}
\usepackage{subdepth}\usepackage{subfig}\usepackage{titlesec}\usepackage[dotinlabels]
{titletoc}\usepackage{wrapfig}\usepackage[all]{xy}\usepackage{young}
\usepackage[vcentermath]{youngtab}\usepackage{relsize}\usepackage{stackengine}
\usepackage{datetime}\usepackage{hyperref}\numberwithin{equation}{section}

\usepackage{verbatim}

   \newcommand{\be}{\begin{equation}}
\newcommand{\ee}{\end{equation}}\newcommand{\bml}{\begin{multline}}\newcommand{\emll}{\end{multline}}
\def\({\left(} \def\){\right)}\def\[{\left[} \def\]{\right]}
\def\Im{\text{Im}}

\def\mE{\mathcal{E}}\def\al{\alpha}\def\mK{\mathcal{K}}
\def\mO{\mathcal{O}}
\def\eps{\epsilon}
\def\v{\vec}\def\mE{\mathcal{E}}
\def\a{\alpha}\def\g{\gamma}\def\lam{\lambda}\newcommand{\G}{\Gamma}\def\d{\partial}\def\o{\omega}\def\XXint#1#2#3{{\setbox0=\hbox{$#1{#2#3}{\int}$} \vcenter{\hbox{$#2#3$}}\kern-.5\wd0}}   \newcommand{\la}{\langle} \newcommand{\ra}{\rangle}\newcommand{\bea}{\begin{eqnarray}}\newcommand{\eea}{\end{eqnarray}}

\def\o{{\omega}}

\usepackage[left=2cm,right=2cm,top=2cm,bottom=2cm]{geometry}
\titleformat{\section}{\normalfont\bfseries}{\thesection.}{4pt}{}\titlespacing{\section}{0pt}{22pt}{6pt}
\titleformat{\subsection}{\normalfont\itshape}{\thesubsection.}{4pt}{}\titlespacing{\subsection}{0pt}{18pt}{6pt}
\titleformat{\subsubsection}{\normalfont\itshape}{\thesubsubsection.}{4pt}{} \titlespacing{\subsubsection}{0pt}{16pt}{6pt}

 \def\ie{\begin{equation}\begin{aligned}} \def\fe{\end{aligned}\end{equation}}

 \def\d{\partial}    \def\1{{\mathds 1}}
 \def\Im{\mathop{\rm Im}}
\def\mN{\mathcal{N}}  \def\mL{\mathcal{L}}
\def\o{\omega}\def\v{\vec }

\DeclareFontShape{OT1}{cmr}{mx}{n}%
    {<->cmr10}{}
\newcommand{\mytitlefont}{\fontseries{mx}\selectfont}
\DeclareMathAlphabet{\titlemath}{OT1}{cmr}{mx}{n}

\newcommand{\bi}{\begin{itemize}}\newcommand{\ei}{\end{itemize}}
\newcommand{\sss}{\subsubsection}

\begin{document}

\begin{titlepage}

\begin{center}

~\\[1cm]

{\fontsize{20pt}{0pt} \mytitlefont
 Degrees of universality in wave turbulence 
}
\\[10pt]

~\\[0.2cm]

{\fontsize{14pt}{0pt}Jiasheng Liu{\small $^{1}$}, Vladimir Rosenhaus{\small $^{1}$}, and Gregory Falkovich{\small $^{2}$}}

~\\[.3cm]
\newdateformat{UKvardate}{%
 \monthname[\THEMONTH]  \THEDAY,  \THEYEAR}
\UKvardate
\today
~\\[0.3cm]

\it{$^1$ Initiative for the Theoretical Sciences}\\ \it{ The Graduate Center, CUNY}\\ \it{
 365 Fifth Ave, New York, NY 10016, USA}\\[.5cm]
 
 \it{$^2$Weizmann Institute of Science}\\ \it{Rehovot 76100 Israel}

~\\[0.6cm]

\end{center}

\noindent

Turbulence of weakly interacting waves displays a great deal of universality: independence of the details of the interaction and of the pumping and dissipation scales. Here we study how inverse turbulent cascades (from small to large scales) transition from weak to strong. We find that while one-loop corrections can be dependent on excitation and dissipation scales, new types of universality appear in strong turbulence.  We contrast turbulence of spin waves in ferromagnets with turbulent cascades in the Nonlinear Schr\"odinger Equation (NSE) and in an MMT-like model in higher dimensions having a multiplicative interaction vertex:  vertex renormalization gives rise to dependence on the pumping (UV scale) in the former but not in the latter. As a result of this spectral nonlocality, spin-wave turbulence stops being weak if one is sufficiently far from the pumping scale, even when the interaction of waves with comparable wavenumbers is weak. We  paraphrase this as:\textit{ nonlocality enhances nonlinearity}. 

We then describe strong turbulence in a multi-component version of these models with a large number of components. We argue that strong spin-wave turbulence is similar to turbulence of the focusing NSE, as it realizes  a critical-balance state. However, UV nonlocality causes the level of spin-wave turbulence at large scales  to \textit{decrease} with increasing  pumping level, culminating in a state that is independent of the level of pumping.

\vfill 
\end{titlepage}

\tableofcontents 

~\\\vspace{.1cm}

\section{Introduction}

When energy is injected into a system at one scale and absorbed at another, one might naturally expect the energy transfer to depend on both the injection and dissipation scales. In this context, the universality of Kolmogorov's four-fifth law of incompressible turbulence is striking: energy flux determined by the third moment of velocity, independent of distant pumping and dissipation scales. This universality, however, is not true for the  higher and lower moments, which acquire anomalous exponents, introducing dependence on the pumping (largest) scale.

Wave turbulence provides a rich arena in which to quantitatively and systematically study spectral redistribution of energy or other conserved quantities. Remarkably, in some cases the direct analog of the four-fifth law can be derived as a universal flux relation of some third or fourth moment. For inverse cascades, such relation was derived for Nonlinear Schr\"odinger Equation  \cite{Nonlocal} and is derived below for spin-wave turbulence. In both cases, one has an exact universal expression for the action flux given by the fourth moment, independent of the pumping and dissipation scales.

Exactly like in Kolmogorov turbulence, the task is now to derive the most physically important quantity -- the second moment -- which gives the spectrum of turbulence in wavenumber space. While this task is so far unsolved for  incompressible turbulence, wave turbulence allows for the limit of small amplitudes in which higher moments can be consistently expressed via the integrals of the second moments. Provided these integrals converge, one obtains the universal Kolmogorov-Zakharov (KZ) spectra of weak turbulence \cite{ZLF}. The spectra are independent of the pumping and damping scales and of the details of the interaction, except for its overall homogeneity degree. In this work, we  explicitly show which of these universality properties are violated at higher nonlinearity for certain classes of models, including  spin-wave turbulence. 

\begin{figure}[h]
\centering
\includegraphics[width=2in]{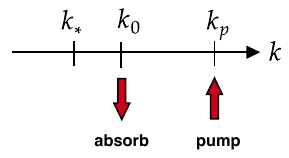}
\end{figure}

For spin waves, weak turbulence turns into strong turbulence for low $k$, so we  consider an inverse cascade of wave action, with pumping at a UV momentum $k_p$ and absorption at an IR momentum $k_0$. Let us denote by $k_*$ the wavenumber where spectrally local interaction ceases to be weak. One remains within the weak turbulence regime for $k_0>k_*$, where the closed description is provided by a kinetic equation on the occupation numbers $n_k$.

At the next-to-leading order in the nonlinearity, we find explicit dependence on the UV and IR cutoffs ($k_p$ and $k_0$, respectively). In particular, that introduces $\mK \equiv k_* \(k_p/k_*\)^{2/3}$ as the lowest momentum at which the  KZ description is valid. Wave turbulence ceases to be weak if one moves sufficiently far from the pumping scale, even at momenta that are  high enough  to make the nonlinear interaction term arbitrarily small. In other words:\textit{ nonlocality enhances nonlinearity}. 

 From the structural form of the higher-order kinetic equation \cite{RS1,RS2, RSSS, Hu:2025bqi}, it follows that the third aspect of KZ universality --- independence of the details of the interaction --- is broken for spin waves. We then consider  another model where the interaction vertex has a different dependence on angles and momenta ratios but the same homogeneity degree  --- and hence  the same KZ state.  In this model we  show that, remarkably,   the next-to-leading order terms have no dependence on $k_0$ and $k_p$. 

The plan of the paper is as follows.
In Sec.~\ref{Sec2} we review the standard kinetic equations for weakly interacting systems and their corresponding inverse cascade solutions in three simplest systems: the nonlinear Schr\"odinger equation (a constant interaction vertex), spin waves (an interaction vertex that is a sum of two term), and a product-factorized vertex model \cite{MMT} picked to have the same KZ state as spin waves. In Sec.~\ref{Sec3} we study the kinetic equation to the next-to-leading order. In subsection~\ref{Sec31}, we review the derivation of the spin wave Hamiltonian. The next-to-leading order kinetic equation is found in subsection~\ref{Sec32}. In subsection~\ref{Sec33} we show how, in the background of the KZ state, the new higher-order term is dependent on the pumping scale and modifies the KZ state of spin waves: The spectrum becomes steeper, and the steepening occurs much earlier (higher $k$) in the cascade than naively expected. In subsection~\ref{Sec35} we show that this dependence on the pumping scale is absent in the NSE and MMT-like models. In subsection~\ref{Sec34} we argue that at low $k$ the effective description for the inverse spin wave cascade should resemble that of an inverse cascade for the focusing nonlinear Schr\"odinger equation. 

In Sec.~\ref{largeNsec} we extend the models of Sec.~\ref{Sec2} to multicomponent fields. In the limit of a large number of components, one derives an integral equation for the renormalized vertex. The solution of this equation and the form of the renormalized vertex were known for NSE. The peculiar form of the exchange interaction allows us to solve this equation analytically for spin waves as well.  We then analyze the new renormalized kinetic equation which is valid at all scales, with  arbitrarily strong nonlinearity. We argue that along the turbulent cascade, from short to long scales, the solution evolves from Kolmogorov-Zakharov to critical balance scaling. While for the former the turbulence level increases with increasing flux, for the latter it decreases until it saturates.

 We conclude in Sec.~\ref{Sec5}. Appendix~\ref{AppOne} provides some technical details relevant to Sec.~\ref{Sec3}. 

\section{Weak wave turbulence at leading order} \label{Sec2}
\begin{figure} \centering
\includegraphics[width=1.2in]{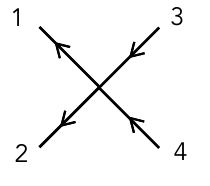}
\caption{The leading order kinetic equation (\ref{KE1}) encodes the tree-level process of two modes directly scattering into two other modes. } \label{Ptree}
\end{figure}

The general Hamiltonian of a system in which waves interact via four-wave scattering is given by 
 \be
 H = \int\! d\v k\, \o_k |a_k|^2+ \int \v d p_1 \v d p_2 \v d p_3 \v d p_4 \,\lam_{1234} a_{1}^{*} a_{2}^{*} a_{3} a_{4}  \delta(\v p_{12;34}) ~,
 \label{Hamiltonian}
 \ee
where $\v p_{12;34}\equiv \v p_1 {+}\v p_2{-}\v p_3 {-}\v p_4$ and $\lam_{1234}$ is a function of the four momenta, $p_1, \ldots, p_4$. The Fourier harmonics $a_i\equiv a_{p_i}$ are normal canonical variables which satisfy the equations of motion $i \frac{\d a_k}{\d t}=\frac{\d  H}{\d  a_k^*}$. The kinetic equation describes the time derivative of the occupation numbers $n_i=\langle |a_{i}|^2\rangle$. Using the equations of motion, this can can be expressed in terms of the four-mode correlation function (fourth cumulant),
\be \label{eomKE}
\frac{\d n_1}{\d t} \equiv I_{p_1}=4\int d\v p_2 d\v p_3 d\v p_4\, \lam_{1234}\, \delta(\v p_{12;34})\Im \la a_1^*a_2^*a_3a_4\ra\,.~
\ee
Scattering conserves the total number of waves, so that we can rewrite the kinetic equation (\ref{eomKE}) as the action continuity equation, $k^d {\partial n_k/ \partial t} =-\partial Q(k)/\partial \v k$. A stationary state of an inverse cascade corresponds to a constant action flux, $Q(k)=Q$:
\be \label{fluxG}
 \int_0^k d \v q\,  I_q=Q~.
\ee
This relation is an analog of the Kolmogorov 4/5-law; it fixes the fourth cumulant. 
The cumulant can,  in turn,  be perturbatively expressed via $n_k$ as a series in $\lambda$ \cite{RS1, RSSS}. At zeroth-order in $\lam$ (no interaction) we assume Gaussian statistics of non-correlated waves, so that the right-hand side of (\ref{eomKE}) vanishes. The first non-vanishing contribution, corresponding to direct two-to-two wave scattering  shown in Fig.~\ref{Ptree}, gives the standard kinetic equation:
\be \label{KE1}
\frac{\d n_1}{\d t} = 16\pi \int d\v p_2 d\v p_3 d\v p_4\,  \lam_{1234}^2 n_1 n_2 n_3 n_4 \Big( \frac{1}{n_1} {+} \frac{1}{n_2}{-}\frac{1}{n_3} {-} \frac{1}{n_4} \Big)  \delta(\o_{12;34})\delta(\v p_{12;34})~,
\ee
where  $\o_{12;34} \equiv\o_{1} {+} \o_{2}{-}\o_{3}{-}\o_{4}=0$ defines the resonant surface. 

Until recently, all analytic studies of wave turbulence were based on the leading order in $\lam$ kinetic equations like  (\ref{KE1}). In order to generate an inverse turbulent cascade solution, one adds forcing and dissipation terms  at high and low $k$, respectively, so that the flux can be emitted and absorbed: terms of the form, e.g., $-\g_k n_k$, are added to the right-hand side, where $\g_k$ is negative for forcing and positive for dissipation. 
The Hamiltonian coefficients are assumed to be scale-invariant:  $\o_k \sim k^{\al}$ and $\lam_{1234} \sim \lam k^{\beta}$. The scaling exponents for the Kolmogorov-Zakharov solutions (\ref{KZe}) for an inverse cascade
\cite{ZLF},
\be \label{KZe}
n_k\sim \(\frac{Q}{\lam^2}\)^{1/3}k^{-\g}~, \ \ \ \g =  
 d+\frac{2}{3}\beta - \frac{\al}{3}
\ee
are simple to establish by  looking at scaling of the collision integrals in the kinetic equation. 
The wave action flux \eqref{fluxG}  scales as $Q\sim \frac{n_k k^d}{\tau_k} \sim \lam^2(n_k k^d)^{3} k^{2\beta - \al}$, where we used the kinetic equation to set the time scales $\tau_k$. A constant $Q$ then gives the occupation numbers (\ref{KZe}). 

For (\ref{KZe}) to be a stationary solution of the kinetic equation requires locality: convergence of the integrals in the kinetic equation upon substituting (\ref{KZe}), which must be checked on a case-by-case basis. The structure of the kinetic equation, having differences of the $n_i$, makes the power of the momentum in the integrand lower by one in the UV. This improved convergence provides a locality window for $\gamma$ in certain cases \cite{ZLF}. In the next section, we will see that higher order terms in the kinetic equation may violate locality, making the occupation numbers $n_k$ dependent on both the UV and IR cutoffs. Of course, the flux law (\ref{fluxG}) expressed in terms of the fourth moment is exact and local: it depends only on the local wavenumber $k$  and is  independent of the UV and IR cutoffs \cite{Nonlocal}. 

Let us now review the KZ states in three specific systems. 

\subsubsection*{Nonlinear Schr\"odinger equation}

The simplest quartic Hamiltonian is the nonlinear Schr\"odinger equation, 
\be
H=\int d^d x \left(|\nabla\psi|^2+\lambda|\psi|^4\right)~, \label{NSE}
\ee
for the complex field $\psi$. It is arguably the most universal model in physics, describing spectrally narrow wave packets and long-wavelength limits of systems of waves and particles, see e.g. \cite{OT,Naz,exp1,exp2,exp3,exp4,Nowak:2011sk,Pom,B15,Zhu_2023, Naz23, zhu2025,Pre,PT,Gaz19}. It also describes Bose-Einstein condensates (under the name of the Gross-Pitaevskii equation). 
The action-flux law \eqref{fluxG} can be written as the condition that the fourth-order two-point cumulant is independent of the distance between  points $1$ and $2$: $2\text{Im}\,\langle \psi^*_1\psi_2|\psi_2|^2\rangle =-Q$ \cite{Nonlocal}. 
Specializing the weak turbulence expression (\ref{KZe}) to this case, which has $\al = 2$ and $\beta = 0$, gives the inverse cascade, 
\be  
 n_k=k^{-d+2/3}(Q/\lambda^2)^{1/3}~.
\ee
The ratio of the interaction energy to the kinetic energy defines the dimensionless nonlinearity parameter, $\epsilon_k=\lambda n_kk^{d-2}\propto \lam k^{-4/3}$. The nonlinearity is strong in the IR, where KZ is expected to break down. 

\subsubsection*{Spin waves}
The next simplest Hamiltonian has two derivatives in the interaction term \cite{Lut, Fujimoto_2016, PhysRevA.108.013318,  PhysRevLett.125.230601, Rodriguez_Nieva_2021, li2023far},
\be
H=\int d^d x \[\o_0|\psi|^2+|\nabla\psi|^2+\lam\(\psi^2(\nabla  \psi^*)^2+\text{c.c.}\)\]~, \label{Ham}
\ee
where $\text{c.c}$ denotes the complex conjugate. 
For $\lam>0$, this describes spin waves in a Heisenberg ferromagnet, provided that the waves have small magnitude. In more detail,   all the spins are aligned in the same direction in an unperturbed state.  Local deviations of the magnetization from the mean lead to precession, which propagates as a spin wave with frequency $\o_k =\omega_0+ k^{2}$, where the constant $\o_0$ will be dropped in what follows. The waves interact via four-wave scattering, described by (\ref{Ham}). In Fourier space (\ref{Hamiltonian}), the interaction is 
\be \label{lam1}
\lam_{1234} = - \lam (\v p_1{\cdot } \v p_2 +\v p_3 {\cdot}\v p_4)~,
\ee
 which changes sign  upon passing through the 3-wave resonance $\omega_{{\v p_1}+{\v p_2}}=|{\v p_1}+{\v p_2}|^2=\omega_1+\omega_2=p_1^2+p_2^2$. This  implies  repulsion of counter-propagating waves and attraction of co-propagating waves. 

The exact equation for the occupation numbers \eqref{eomKE} now takes the form
\be \label{eomKEs}
\frac{\d n_1}{\d t} \equiv I_{p_1}=-4\lam\int d\v p_2 d\v p_3 d\v p_4\, (\v p_1{\cdot } \v p_2 +\v p_3 {\cdot}\v p_4)\, \delta(\v p_{12;34})\Im \la \psi_1^*\psi_2^*\psi_3\psi_4\ra\,.~
\ee
It conserves the  wave action  $\int n_k\,d\v k$, which in this case is the difference between the total magnetization and its $z-$component, which commutes with the Hamiltonian. We can then write  (\ref{eomKEs}) as the action continuity equation, $ {\partial n_k/ \partial t} =-\nabla_{\v k} \,Q_k $. The constant-flux condition \eqref{fluxG} now takes the form:
\be \int_0^qk^{d-1}dk d\Omega \,I_k=Q=\Im\left\langle[ \psi_1\psi_2^*(\nabla\psi_2)^2 -\psi_1\nabla\psi_2^2\nabla\psi_2^*)\right\rangle\ .\label{flux}\ee
The flux is independent of wavenumber $q$, or equivalently, is independent of the distance $r_{12}$ between the points in the fourth-order two-point cumulant (the right-hand side of (\ref{fluxG}), which is an analog of the Kolmogorov's 4/5 law).

In the weak-turbulence limit, the interactions relevant to the kinetic equation (\ref{KE1}) are localized on the resonant surface, $\o_1{+}\o_2=\o_3{+}\o_4$, which may equivalently be expressed as $\v p_1{\cdot} \v p_2= \v p_3{\cdot } \v p_4$. Since the constant $\o_0$ in the dispersion relation makes no appearance in the kinetic equation, we drop it. 

Specializing (\ref{KZe}) to this case, which has $\al = 2$ and $\beta = 2$, gives the KZ  inverse cascade, 
\be  \label{KZspin}
n_k\simeq Q^{1/3}\lambda^{-2/3}k^{-d-2/3} ~.
\ee
The expectation --- which we will see is too naive --- is that the KZ state is valid as long as the dimensionless nonlinearity parameter $\eps_k$ is small. It takes the form, 
\be \label{39}
\epsilon_k \sim \lam n_k k^d\sim (Q\lambda)^{1/3} k^{-2/3}\sim \(\frac{k_*}{k}\)^{2/3}~,
\ee
where we defined $k_* \equiv \sqrt{Q\lam}$ at which $\eps_k$  becomes equal to one. 
Like for the nonlinear Schr\"odinger equation, $\eps_k$ grows with decreasing $k$. Physically, $\eps_k$ is the precession angle of the magnetic moment in a spin wave.~\footnote{In a direct cascade, which has KZ scaling $n_k \sim k^{-d -4/3}$, the nonlinearity parameter $\eps_k \sim k^{-4/3}$ decreases along the cascade. For this reason, we study the transition to strong turbulence in an inverse cascade. }

\subsubsection*{Factorized interaction model}
It is instructive to compare the  above two cases with that of a  Hamiltonian having a  factorized  interaction:  $\lam_{1234} =\lam  (p_1 p_2 p_3 p_4)^{1/2}$. This kind of interaction appears in the one-dimensional MMT model \cite{MMT, MMT2, MMT3, MMT4, MMT6, MMT5}, though we will work in arbitrary spatial dimension. The interaction is not analytic, so it is not particularly natural unless it arises from integrating out other fields; in position space it takes the form, 
\be \label{MMT}
H=\int d^d x \(|\nabla\psi|^2+ \lam |(\nabla^2)^{1/4} \psi |^{4}\)~.
\ee
The point, however, is that just like the spin wave Hamiltonian, this Hamiltonian has the scaling exponents $\al = \beta = 2$. Therefore,  for weak turbulence, it has the exact same $n_k$  (\ref{KZspin}) as spin waves. We will see in the next section that, in contrast,  the next-to-leading order corrections to $n_k$ differ drastically for these two Hamiltonians.

\section{Spin wave turbulence at next-to-leading order } \label{Sec3}

\subsection{Spin Waves} \label{Sec31}
In the previous section we wrote down a Hamiltonian with a two-derivative interaction (\ref{Ham}). This naturally arises in the description of a Heisenberg ferromagnetic, as we now review.

 The Heisenberg ferromagnet is described by the Hamiltonian, 
\be \label{A1}
H =\frac{1}{8\lam} \int d^d x\,  (\nabla \vec S)^2~\ \ \ \ \ \  \ \ \vec S^2 = 1~.
\ee
The spin vector has three components, $\v S= (S_1, S_2, S_3)$, which are not independent due to the constraint $\vec S^2 = 1$. One can eliminate   $S_3 = \sqrt{1{-} S_1^2 {-} S_2^2}$, giving a Hamiltonian in term of the first two components
\be
H =\frac{1}{8\lam} \int d^d x \( (\nabla \vec S)^2 + \frac{(\v S {\cdot} \nabla \v S)^2}{1- \v S^2}   \)~, \ \ \ \v S= (S_1,S_2)~,
\ee
where: $(\v S{\cdot} \nabla \v S)^2 \equiv S_a \d_i S_a S_b \d_i S_b$, the indices $a,b$ run from $1$ to $2$, $i$ runs from $1$ to $d$, and repeated indices are summed over. 
 The spins transform in the fundamental representation of $SO(3)$, with corresponding Poisson brackets, 
\be \label{com}
\{S_i, S_j\} = \eps_{i jk} S_k~,
\ee
where $\eps_{i j k}$ is the anti-symmetric Levi-Civita symbol. The equation of motion is the Landau-Lifshitz equation, 
\be
\frac{\d S_i}{\d t} = \{S_i, H\} =\frac{1}{4\lam} \eps_{i j k} S_j \nabla^2 S_k~.
\ee
In one dimension this model is integrable, and can in fact be transformed into the nonlinear Schr\"odinger equation; we will be in higher dimensions, generally $d=3$. 

The noncanonical algebra (\ref{com}) of the spin variables makes them awkward to work with. Perturbatively, we may transform to canonical variables: we exchange the two independent real fields $S_1$ and $S_2$ for a complex field $\psi$, 
\be
\frac{1}{2\sqrt{2\lam} }(S_1 +i S_2 )\equiv \psi \sqrt{1 - 2\lam |\psi|^2}~, \ \ \ \ S_3 = 1- 4\lam |\psi|^2~,
\ee
making it manifest that $\vec S^2  =|S_1 {+}i S_2|^2 + S_3^2  = 1$. We assume that the spins are close to being aligned in the $z$ direction, $S_3 \lessapprox 1$, so that $\psi$ is small. This is the regime in which small perturbations propagate as spin waves. Expanding the Hamiltonian (\ref{A1}) perturbatively in $\psi$ up to sixth order, yields 
\be \label{Hspin2}
H\approx \int d^d x\, \Big( |\nabla \psi|^2 +\lam(\psi^2 (\nabla \psi^*)^2 + \text{c.c}) + \lam^2\((\psi^2 (\nabla \psi^*)^2 + \text{c.c})|\psi|^2 + 2|\nabla \psi|^2 |\psi|^4  \) +\ldots \Big)~.
\ee
At linear order in $\lam$, this reproduces the above Hamiltonian  (\ref{Ham}) and gives the order $\lam^2$ kinetic equation (\ref{KE1}) with interaction (\ref{lam1}). 

\subsection{Next-to-leading order kinetic equation} \label{Sec32}

Our goal now is to study the kinetic equation for (\ref{Hspin2}) up to order $\lam^3$. We may write it as, 
\be \label{KEF}
\frac{\d n_k}{\d t} = \frac{\d n_k}{\d t}\Big|_{\mO(\lam^2)} + \frac{\d n_k}{\d t}\Big|^{\text{sextic}}_{\mO(\lam^3)} +\frac{\d n_k}{\d t}\Big|^{\text{1-loop}}_{\mO(\lam^3)} + \ldots~.
\ee
The first term is the $\lam^2$-term from before. At order $\lam^3$ 
 there are two qualitatively different contributions:  the second term on the right in (\ref{KEF}) which arises from the sextic interaction term in (\ref{Hspin2}) and  is computed in Sec.~\ref{sextic}, and the third term on the right of (\ref{KEF})  which is due to higher order contributions of the quartic interaction in (\ref{Hspin2}) and is computed in Sec.~\ref{oneloop}. 

\subsubsection{Sextic interactions} \label{sextic}
\begin{figure}
\centering
\includegraphics[width=.9in]{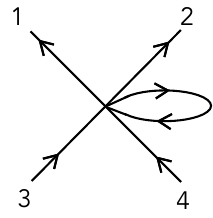}
\caption{The contribution of the sextic interaction to the fourth cumulant.}  \label{Goneloop}
\end{figure}
Since the Hamiltonian (\ref{Hspin2}) has both quartic and sextic interactions, the kinetic equation (\ref{eomKE}) that was written for theories with only quartic interactions needs to be modified to account for the sextic term. This gives, 
\be 
\!\!\! \frac{\d n_1}{\d t}=4\int \!\prod_{i=2}^4 d \v p_i\,  \lam_{1234}\, \delta(\v p_{12;34})\Im \la a_1^*a_2^*a_3a_4\ra + 6\!\int\! \prod_{i=2}^6\,  d \v p_i  \lam_{123456}\, \delta(\v p_{123;456})\Im \la a_1^*a_2^*a_3^*a_4 a_5 a_6\ra~. 
\ee
Since the sextic interaction scales as $\lam^2$, and the sixth cumulant is also of order $\lam^2$, the second term in this kinetic equation is of order $\lam^4$, and so can be neglected at the order $\lam^3$ to which we will be  working. 

Thus, we only need the leading contribution of the sextic terms in the Hamiltonian to the fourth-cumulant. This is trivial to get, by simply contracting two of the $\psi$'s, see Fig.~\ref{Goneloop}. This yields an effective quartic interaction, 
\be \label{H64}
H_{6\rightarrow 4}=\lam^2 \int d^d x\, \Big((\psi^2 (\nabla \psi^*)^2 + \text{c.c})\, \mN + 8|\nabla \psi|^2 |\psi|^2 \mN + 2\mathcal{E}|\psi|^4  \Big)~,
\ee
where we defined $\mN$ and $\mE$ as the total wave action and kinetic energy, respectively, 
\bea \label{mN}
\mN &\equiv& \la |\psi|^2\ra = \int d \v k\,  n_k \\ \label{mE}
\mE& \equiv& \la |\nabla \psi|^2\ra = \int d \v k\, k^2 n_k~. 
\eea
 Fourier transforming $H_{6\rightarrow 4}$,  we may write the interaction in the form (\ref{Hamiltonian}) with,
\be
\lam_{1234}\Big|_{6\rightarrow 4} = \lam^2 \Big(- \mN (\v p_1{\cdot} \v p_2{+}\v p_3 {\cdot} \v p_4)+ 2 \mN( \v p_1{+}\v p_2){\cdot}(\v p_3 {+}\v p_4)+2 \mE \Big)~.
\ee
Therefore, the fourth cumulant that we insert into the kinetic equation is just the tree-level one with this interaction. So in (\ref{KE1}) we simply need to replace $\lam_{1234}^2$ with $\lam_{1234}\,  \lam_{1234}\Big|_{6\rightarrow 4} $, giving, 
\bml \label{313}
\frac{\d n_1}{\d t}\Big|^{\text{sextic}}_{\mO(\lam^3)}  =  16\pi \lam^3 \int d\v p_2 d\v p_3 d\v p_4\,
 n_1 n_2 n_3 n_4 \Big( \frac{1}{n_1} {+} \frac{1}{n_2}{-}\frac{1}{n_3} {-} \frac{1}{n_4} \Big)  \delta(\o_{12;34})\delta(\v p_{12;34})\\ 
  (\v p_{1}{\cdot }\v p_2 + \v p_3{\cdot } \v p_4) \Big(   (\v p_{1}{\cdot }\v p_2 + \v p_3{\cdot } \v p_4)  \mN - 2  \mN( \v p_1{+}\v p_2){\cdot}(\v p_3 {+}\v p_4)-2 \mE \Big)~.
 \end{multline}
 We may use the momentum and frequency conserving delta functions to simplify this to, 
 \bml \label{314}
 \frac{\d n_1}{\d t}\Big|^{\text{sextic}}_{\mO(\lam^3)}  =  64\pi \lam^3 \int d\v p_2 d\v p_3 d\v p_4\,
 n_1 n_2 n_3 n_4 \Big( \frac{1}{n_1} {+} \frac{1}{n_2}{-}\frac{1}{n_3} {-} \frac{1}{n_4} \Big)  \delta(\o_{12;34})\delta(\v p_{12;34})\\ 
  (\v p_{1}{\cdot }\v p_2)^2  \Big(    \mN -   \frac{\mN( \v p_1{+}\v p_2)^2}{\v p_{1}{\cdot }\v p_2}-\frac{ \mE }{\v p_{1}{\cdot }\v p_2}\Big)~.
 \end{multline}

If it is simply the kinetic equation that one is interested in, one should insert (\ref{314}) into (\ref{KEF}) with $\mN$ and $\mE$  defined in (\ref{mN}) and (\ref{mE}) as integrals of $n_k$. However, we are  particularly interested in the kinetic equation in the vicinity of the KZ state, in order to see how the higher order terms impact the form of the stationary state and take us away from the KZ state. In this case, the $\mN$ and $\mE$ that appear in (\ref{314}) should be evaluated in the KZ state. 
For the KZ state for spin waves we find that $\mN$ depends on the IR cutoff (the momentum $k_0$ at which the KZ spectrum is terminated, either by a sink or by transitioning into a flatter spectrum of strong turbulence, as described below). The kinetic energy $\mE$ depends on the UV cutoff (the momentum $k_p$ at which flux is pumped), 
\bea \label{mN2}
\mE &=&
  \int^{k_p} \frac{d^d k}{(2\pi)^d} k^2\, n_k  = \frac{3}{4}S_d \lam^{-2/3} Q^{1/3} k_p^{4/3}~, \ \  \ \ \ \ S_d = \frac{2}{(4\pi)^{d/2} \G(d/2)}
\\ \label{mE2}
\mN& =& \int_{k_0} \frac{d^d k}{(2\pi)^d}\, n_k = \frac{3}{2}S_d \lam^{-2/3}  Q^{1/3}k_0^{-2/3}~.
\eea

 \subsubsection{One-loop diagrams for quartic interactions} \label{oneloop}
 \begin{figure}
\centering 
\subfloat[]{\includegraphics[width=1.8in]{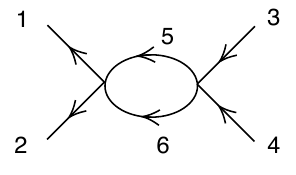}} \ \ \ \ \ \   \ \ \ \ \ \ 
\subfloat[]{\includegraphics[width=1.8in]{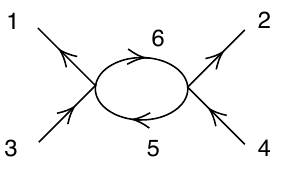}}
\caption{Feynman diagrams for (a) $\mL_+$ and (b) $\mL_-$~.} \label{LoopFig}
\end{figure}

Now let us look at the one-loop contributions for the sector of the Hamiltonian with quartic interactions. 
For a general theory with a quartic interaction (\ref{Hamiltonian}), the kinetic equation to order $\lam^3$ is given by \cite{RS1}
\be     \label{onelevelKE}
    \frac{\d n_1}{\d t} = 16\pi\int d \vec{p}_2 d \vec{p}_3 d \vec{p}_4 \lambda_{1234}^2 n_1 n_2 n_3 n_4\Big(\frac{1}{n_1}{+}\frac{1}{n_2}{-}\frac{1}{n_3}{-}\frac{1}{n_4}\Big)
    \Big(1 + 2\mL_+ + 8\mL_- \Big)\delta(\o_{12;34}) \delta(\v p_{12;34})~.
\ee
where 
\be \label{Lpm}
\mL_+ = 2\!\int\!\! d \vec{p}_5 d \vec{p}_6 \frac{\lam_{1256}\lam_{3456}}{\lam_{1234}} \frac{n_5{+}n_6}{\o_{12;56}  }\delta(\v p_{12;56})~, \ \ \ \ \mL_-= 2\!\int\!\! d\vec{ p}_5 d\vec{ p}_6  \frac{\lam_{3516}\lam_{4625}}{\lam_{1234}} \frac{n_6{-}n_5}{\o_{16;35} }\delta(\v p_{16;35})~,
\ee
where the $1/\omega$ denominators are understood as their principal values.~\footnote{Our conventions are $d\v k \equiv \frac{d^d k}{(2\pi)^d}$, and correspondingly when actually evaluating integrals $\delta(\v p_{12;34})$ should be replaced with $(2\pi)^d \delta(\v p_{12;34})$.} The loop integrals are illustrated in Fig.~\ref{LoopFig}.

The quartic interaction $\lam_{1234}$ in our case is given by (\ref{lam1}), which should be inserted into (\ref{Lpm}). The expression can be simplified slightly, as is done in Appendix~\ref{AppOne}, see (\ref{A4}). 
For the purposes of perturbatively finding the stationary solutions, we should evaluate $\mL_{\pm}$ for an $n_k$ that is given by the KZ state. This computation is done in Appendix~\ref{AppOne} with the result, 
 \be \label{oneLQ}
\frac{\d n_1}{\d t}\Big|^{\text{1-loop}}_{\mO(\lam^3)}  = 64\pi \lam^2 \int d\v p_2 d\v p_3 d\v p_4\,
 n_1 n_2 n_3 n_4 \Big( \frac{1}{n_1} {+} \frac{1}{n_2}{-}\frac{1}{n_3} {-} \frac{1}{n_4} \Big)  \delta(\o_{12;34})\delta(\v p_{12;34}) (\v p_1{\cdot} \v p_2)^2 \mL~,
\ee
where  $\mL$ is given by, 
\be \label{324}
\mL = 2\lam \mN + 2  \lam\frac{(\v p_1{-}\v p_3)^2}{\v p_1{\cdot} \v p_2} \mN+2 \frac{\lam}{\v p_1{\cdot }\v p_2} \mE + \text{finite}~,
\ee
where $\text{``finite''}$ denotes terms that depend on neither the UV nor IR cutoffs. \\

We may combine the three contributions to the kinetic equation (\ref{KEF}),  with the order $\lam^2$ term given by (\ref{KE1}) and (\ref{lam1}) and the order $\lam^3$ terms given by (\ref{313}) and (\ref{oneLQ}), to write the kinetic equation for spin waves to order $\lam^3$, in the vicinity of the KZ state, 
\bml \label{KEN}
\frac{\d n_1}{\d t} = 64\lam^2 \pi \int d\v p_2 d\v p_3 d\v p_4\, (\v p_1{\cdot} \v p_2)^2 n_1 n_2 n_3 n_4 \Big( \frac{1}{n_1} {+} \frac{1}{n_2}{-}\frac{1}{n_3} {-} \frac{1}{n_4} \Big)  \delta(\o_{12;34})\delta(\v p_{12;34}) \\
\( 1+  \lam \Big(    3 \mN -   \mN \frac{( \v p_1{+}\v p_2)^2}{\v p_{1}{\cdot }\v p_2} +   \frac{2(\v p_1{-}\v p_3)^2}{\v p_1{\cdot} \v p_2} \mN + \frac{\mE}{\v p_1 {\cdot }\v p_2} + \text{finite} \Big) \)~.
\end{multline}

\subsection{Impact of higher order kinetic equation on the turbulent spectrum} \label{Sec33}
 
Let us  see what impact the additional (order $\lam^3$ terms) in the next-to-leading order kinetic equation (\ref{KEN})  has on the KZ state. The first term simply rescales the interaction, $\lam^2\rightarrow \lam^2(1+3 \lam \mN)$, and so only affects the prefactor in the KZ state (\ref{KZspin}). This property can be traced to first term in $H_{6\rightarrow 4}$ (\ref{H64}), which simply rescales the interaction term. The second and third terms in (\ref{KEN}) have the same scaling exponent (under a rescaling of all momenta) as the interaction in the leading order kinetic equation, i.e. the interaction continues to have $\beta=2$, so they also don't affect the scaling exponent of $n_k$. This result can already be seen from how the corresponding term appears in $H_{6\rightarrow 4}$: the second term $8 \lam^2 \mN |\nabla \psi|^2 |\psi|^2$ may be absorbed into the kinetic term $|\nabla \psi|^2$ through a field redefinition.~\footnote{Explicitly, sending $\psi \rightarrow \psi( 1-2\mN \lam^2  |\psi|^2)$ transforms $|\nabla \psi|^2 \rightarrow |\nabla \psi|^2  - 2\mN \lam^2 \Big(4 |\psi|^2 |\nabla \psi|^2 + (\psi^2 (\nabla \psi^*)^2 +\text{c.c.})\Big) + \mO(\lam^4)$, and so cancels off  the $8 \lam^2 \mN |\nabla \psi|^2 |\psi|^2$ term, at the expense of  changing the coefficient of the first term in (\ref{H64}) and generating high order in $\lam$ terms.}

Finally, the last term in (\ref{KEN}), $ \lam \mE/{\v p_1{\cdot} \v p_2}$, enhances or suppresses the interaction of co- or counter-propagating waves, respectively. Because it scales differently, this term can modify the spectral slope. It also leads to perhaps the most surprising prediction of the next-to-leading-order kinetic equation for spin waves: the explicit appearance of the kinetic energy $\mE$ (\ref{mE}), which, in the KZ state, is sensitive to the UV cutoff (the pumping scale $k_p$). Conventional wisdom states that the strength of the nonlinearity $\eps_k$ --- the ratio of the quartic to quadratic terms in the Hamiltonian, estimated in the KZ state ---  sets the range of validity of weak wave turbulence, meaning for which $k$ it is justified to use the leading-order kinetic equation (\ref{KE1}). For spin waves, the naive nonlinearity parameter $\eps_k=(k_*/k)^{2/3}$  is small as long as $k\gg k_* \equiv \sqrt{Q\lam}$. Therefore, one may have thought that as long as $k_p>k_0\gg k_*$, so that the cascade is terminated at a $k_0$ well within the regime of weak nonlinearity, one can trust (\ref{KE1}). Evidently, this is false:  the next-to-leading order term in (\ref{KEN}) being small requires
\be \label{WWC}
\frac{\lam \mE}{k^2}  \sim \frac{(\lam Q)^{1/3} k_p^{4/3}}{k^2} \ll 1~, \ \ \ \ \Rightarrow k> (\lam Q)^{1/6} k_p^{2/3}\sim k_*\(\frac{k_p}{k_*}\)^{2/3}\equiv \mathcal{K}~, \ \ \ \  k_* \equiv \sqrt{Q\lam}
\ee
which is parametrically larger than $k_*$, and dependent on $k_p$. 
The regime of weak  turbulence requires the $k_p$-dependent condition
\be
k_0 \gg \mK~.
\ee
The dependence of the one-loop correction on the pumping scale (\textit{nonlocality}) causes the transition away from KZ (\textit{nonlinearity}) to happen earlier along the cascade, at $\mK$ instead of $k_*$: \textit{nonlocality enhances nonlinearity in wave turbulence}.

We note that here we considered the spin wave Hamiltonian (\ref{A1}), perturbatively expanded to sextic order (\ref{Hspin2}). If we instead consider a model with only the quartic interaction (\ref{Ham}), the kinetic equation at next-to-leading order will be (\ref{KEF}) without the sextic term contribution. Since the $\mE$ dependent term coming from quartic contribution  (\ref{324}) is twice the value of the  $\mE$ dependent term coming from sextic contribution  (\ref{313}) -- which comes with a relative minus sign -- including or not including the sextic term doesn't change the qualitative result that the $\mE$ contribution comes with a positive sign and makes the spectrum steeper. It would have been intriguing if the $\mE$ terms coming from the quartic and sextic interactions canceled each other, thereby  eliminating the dependence on the UV cutoff ---  but this is not the case.

\subsection{UV behavior} \label{Sec35}

The universality of the KZ solution -- independent of the details of the interaction -- is a consequence of the scale invariance of all the terms of the leading order kinetic equation. The next-order corrections, $\mL_{\pm}$ appearing in (\ref{Lpm}), 
break that scale invariance. As a result, the details of the interaction $\lam_{1234}$ will matter. 

Let us look at the integrand of $\mL_+$ in the UV,  at large $p_5\approx p_6 \gg p_1, \ldots p_4$. From $\lam_{1256} =  - \lam(p_1{\cdot} p_2 + p_5{\cdot} p_6)$, it is clear that dominant contribution is the cross term in $\lam_{5612}\lam_{3456}$ involving $p_5{\cdot} p_6$ from $\lam_{1256}$ and $p_5{\cdot} p_6$ from $\lam_{5634}$, 
\be \label{322}
\mL_+  \sim \lam \int^{k_p}d\v p_5  \, p_5^4\,  \frac{n_5 }{p_5^2}~,
\ee
which is indeed dominated by the UV for the KZ state. 

On the other hand, for the factorized interaction (\ref{MMT}), $\lam_{1256} = \lam (p_1 p_2 p_5 p_6)^{1/2}$, the scaling of $\mL_+$ in the UV is, 
\be
\mL_+  \sim \lam \int^{k_p} d \v p_5 \, p_5^2\,  \frac{n_5 }{p_5^2}~,
\ee
which is down by two powers of $p_5$ relative to (\ref{322}); the integral converges in the UV, and there is no dependence on $k_p$.~\footnote{Here we are discussing convergence of the loop integrals appearing in the kinetic equation, and noting the distinction between the spin waves with an additive vertex and the model with a factorized vertex. As discussed in \cite{FR}, even if the loop integrals converge, the integrals over the momenta in the kinetic equation, $\v p_2, \v p_3, \v p_4$, may have UV cutoff dependence.}  Convergent one-loop integrals is also what we saw in the case of the nonlinear Schr\"odinger equation \cite{RF2}.  

One may ask if the result of the one-loop corrections $\mL_{\pm}$ (\ref{Lpm}) is to make the spectrum steeper or less steep for this product factorized interaction. Addressing this would require further analysis;  although, in the case that the interaction is of this form but with a different scaling,  $\lam_{1234} = \lam (p_1 p_2 p_3 p_4)^{\beta/4}$ with $\beta = 2(2 - \eps)$ with $\eps \ll 1$, it was shown in \cite{Rosenhaus:2025bgy} that the spectrum becomes steeper in the defocusing case,~\footnote{The argument is simple: the essential point is that with small $\eps$, the loop integrals are dominated by the UV. } the same as  for the nonlinear Schr\"odinger equation \cite{RF2, Rosenhaus:2025tjx}. 

\subsection{IR behavior} \label{Sec34}
The change of behavior along the inverse cascade towards smaller wavenumbers can be understood from the perspective of renormalization group (RG) flows \cite{Rosenhaus:2025bgy}. Suppose that at the pumping scale $k_p$ we start off with just the Hamiltonian (\ref{Ham}) having a two-derivative interaction. As we flow into the IR, we start  needing to account for loop diagrams. At the one-loop level we get an ``effective'' Hamiltonian,
 \be \label{Hintf}
 H_{int}(\mu) = \int d^d x \( a(\mu)|\nabla \psi|^4 +  b(\mu) \(\psi^2(\nabla  \psi^*)^2+\text{c.c.} \) +c(\mu) |\psi|^4 \)~,
 \ee
which has, in addition to the two-derivative term in (\ref{Ham}), a new four-derivative term and a zero-derivative term. The form of (\ref{Hintf}) is simple to see: it is a consequence of the one-loop diagram contributions to the fourth cumulant having a product of two interactions, such as $\lam_{1256} \lam_{56 34}$, where $p_5$ and $p_6$ are integrated over. With $\lam_{1234} = - \lam(\v p_1{\cdot} \v p_2 {+}\v p_3{\cdot} \v p_4)$, the product of two interactions can produce: $\v p_1{\cdot} \v p_2\, \v p_3{\cdot} \v p_4$, $\v p_1{\cdot} \v p_2 +\v p_3{\cdot} \v p_4$, and $1$, which are the first, second, and third terms in (\ref{Hintf}), respectively. Producing the third term means that for both of the interaction  vertices a $\v p_5{\cdot} \v p_6$ was chosen, giving a factor of $\mE$ (\ref{mE}). 

In the IR it is the last term  in (\ref{Hintf}) -- with zero-derivatives -- that dominates, as the higher derivative terms are suppressed by the extra powers of small wavenumbers. The one-loop calculation above shows that this nonlinear Schr\"odinger equation term corresponds to attraction.~\footnote{Indeed, the sextic contribution $\mE |\psi|^4$ in (\ref{H64}) is positive (repulsion), but the quartic next-order contribution has the opposite sign and is twice as large.} 
An argument to see this universal IR behavior from RG is that spin waves have modulational instability, a property shared with the focusing nonlinear Schr\"odinger equation. Perturbing around a constant solution gives the behavior $\exp(iqx-\Omega t)$ with $\Omega^2=-\lambda k_0^2q^2|\psi|^2+q^4$. This  describes the modulational instability, leading to wave collapse (self-focusing). In the opposite limit,  perturbations of  constant magnetization have the spectrum $\Omega^2=q^4\left(1-\lambda^2|\psi|^4\right)$, which are also unstable when the amplitude exceeds the threshold. Therefore, we expect  strong turbulence to be qualitatively similar to that in optical turbulence in focusing media,  and be determined by a critical balance between linear and nonlinear effects \cite{Rosenhaus:2025tjx}.

We might summarize this by saying that a $k$-independent attraction of long waves is generated by the interaction mediated by short waves. If in the IR the spin wave system behaves like the nonlinear Schr\"odinger equation, it is reasonable to expect that their strong turbulence behavior is similar --- that the strong turbulence of spin waves is qualitatively similar to that of optical turbulence in focusing media. The latter may be determined by a critical balance between linear and nonlinear effects \cite{Rosenhaus:2025tjx, RF2}. We show that this is the case in the next section.

\section{Strong wave turbulence in large $N$ models} \label{largeNsec}

In order to analytically study wave kinetics at strong nonlinearity, we employ the large $N$ expansion, extending the theories to have $N \gg 1$ fields and solving them to all orders in the coupling $\lam$, at leading nontrivial order in $1/N$. Whereas when discussing weak wave turbulence in the previous section we had to take the momentum scale $k_0$ at which the flux is absorbed to be sufficiently high (so as to always remain in the weak nonlinearity regime), here we may take $k_0\rightarrow 0$. 

\subsection{Large $N$ kinetic equations}
 Concretely, instead of a theory of a single complex field $\psi$, we consider a theory of   a vector of complex fields, $\v \psi$, having $N$ components.
 In Sec.~\ref{Sec2} we looked at three Hamiltonians: the nonlinear Schr\"odinger equation (\ref{NSE}), which we extend to,
 \be
H=\int d^d x \left(|\nabla\v\psi|^2+\frac{\lambda}{N} (\v \psi {\cdot} \v \psi^*)^2\right) \label{NSEN}~,
\ee
the quartic spin wave theory (\ref{Ham}) which we extend to,
\be \label{NHam}
H = \int d^d x \( |\nabla \v \psi|^2 +  \frac{\lam}{N} (\v \psi{\cdot} \v \psi (\nabla  \v \psi^* {\cdot} \nabla \v \psi^*) +\text{c.c} )  \)~,
\ee
and the factorized interaction (\ref{MMT}) which we extend to, 
\be \label{MMTN}
H=\int d^d x \(|\nabla\v\psi|^2+ \lam ((\nabla^2)^{1/4} \v \psi{\cdot}  (\nabla^2)^{1/4}\v \psi^* )^{2}\)~.
\ee
 \begin{figure}[t]
{
\centering  \ \ \  \ \ \ 
\subfloat[]{\includegraphics[width=2.7in]{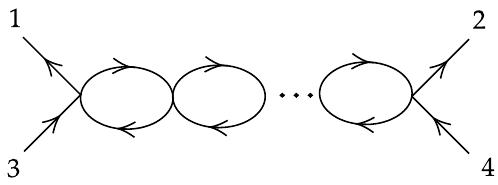}} \ \ \ \ \ \  \ \ \ \ \ \ 
\subfloat[]{\includegraphics[width=2.7in]{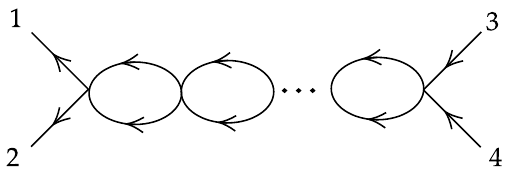}}
}
\caption{The large $N$ kinetic equation is found by summing bubble diagrams. } \label{Nfig}
\end{figure}
\indent The large $N$ kinetic equations for these theories are,  
\be  \label{NKin}
\frac{\d n_1}{\d t} = \frac{8\pi}{N} \int d\v p_2 d\v p_3 d\v p_4\,  |\Lambda_{1234}|^2 n_1 n_2 n_3 n_4 \Big( \frac{1}{n_1} {+} \frac{1}{n_2}{-}\frac{1}{n_3} {-} \frac{1}{n_4} \Big)  \delta(\o_{12;34})\delta(\v p_{12;34})~,
\ee
where: for the large $N$ nonlinear Schr\"odinger (\ref{NSEN}) \cite{bergesGasenzerScheppach2010, berges2002,bergesRothkopfSchmidt2008, Walz:2017ffj, Nowak:2011sk,B15, RF2, RSch},
\be \label{45}
\Lambda_{1234} = \frac{\lam}{1- \mL_-}~, \ \ \ \ \mL_- =   2\lam\int\!\! d\vec{ p}_5 d\vec{ p}_6   \frac{n_6{-}n_5}{\o_{16;35} {+}i\eps}\delta(\v p_{16;35})~,
\ee
for the large $N$ factorized model (\ref{MMTN}) \cite{RSch2}
\be \label{46}
\Lambda_{1234} = \frac{\lam}{1- \mL_-}~, \ \ \ \ \mL_- =   2\lam\int\!\! d\vec{ p}_5 d\vec{ p}_6\,  p_5 p_6 \frac{n_6{-}n_5}{\o_{16;35}{+}i\eps }\delta(\v p_{16;35})~,
\ee
and for the large $N$ quartic spin wave theory (\ref{NHam}) \cite{RSch2}, 
\be \label{Lambda1234}
 \Lambda_{1234} =  \frac{\lam_{12} \lam_{34} \mL_0 + \lam_{1234} (1{-}\mL_1) + \mL_2}{(1{-} \mL_1)^2 - \mL_0 \mL_2}~, \ \ \  \mL_r = 4\!\int\! \!d \v p_5 d\v p_6\, \lam_{56}^r \frac{n_5}{\o_{12;56}{+}i\eps}\delta(\v p_{12;56})~,
 \ee
 where $\lam_{ij} \equiv -\lam \v p_i{\cdot} \v p_j$ and $\lam_{1234} = \lam_{12} + \lam_{34}$ is the original interaction.

These kinetic equations are obtained by summing an infinite number of bubble diagrams. Due to the structure of the dot products, for (\ref{NSEN}) and (\ref{NHam}) the bubbles have arrows going in the opposite direction, see Fig.~\ref{Nfig}(a), while for (\ref{MMTN}) the bubbles have arrows going in same direction, see Fig.~\ref{Nfig}(b). All other Feynman diagrams that aren't these are suppressed by additional factors of $1/N$. 

For the nonlinear Schr\"odinger equation and the product-factorized models, the loop diagrams don't generate any new kinds of interactions \cite{FR}. For the quartic spin wave theory, which has a ``sum factorized'' interaction, the loop diagrams  do generate new kinds of interactions, which are precisely the kind discussed in the context of one-loop renormalization (\ref{Hintf}). 

It is easy to take the weak coupling limit of these kinetic equations. For the large $N$ nonlinear Schr\"odinger equation and the model, (\ref{45}) and (\ref{46}), Taylor expanding the denominator in $\Lambda_{1234}$ gives, $\Lambda_{1234} = \lam (1+ \mL_- +\ldots)$, which is the one-loop answer, in which only the diagram with arrows in opposite directions is kept. 
For the large $N$ quartic spin wave theory, Taylor expanding $ |\Lambda_{1234} |^2$ in  (\ref{Lambda1234}) to order $\lam^3$ gives, 
\be \label{48}
 |\Lambda_{1234} |^2 = \lam_{1234}^2 \( 1+ 2 \text{Re} \(\frac{\lam_{12}\lam_{34}}{\lam_{1234}} \mL_0+ \mL_1 + \frac{\mL_2}{\lam_{1234}}\)\)  =  \lam_{1234}^2 (1 + 2\mL_+)~,
 \ee
 which matches the one-loop kinetic equation (\ref{onelevelKE}) if only the diagram with arrows in the same direction (the $\mL_+$ term) is included, as must be the case.

The large $N$ kinetic equations are valid for any coupling $\lam$, any flux $Q$, and any pumping scale $k_p$ and absorption scale $k_0$. In particular, they allow us to study strong turbulence, which occurs at low wavenumber $k$. This is what we do now for the large $N$ quartic spin wave theory. 
\subsection{Strong turbulence} 

The constant-flux condition can be written schematically as 
\be \label{fluxQ}
Q\simeq \Lambda^2n_k^3k^{3d-2}~.
\ee
 Indeed,  for weak turbulence with $\Lambda_{1234}=\lambda_{1234}\simeq \lambda k^2$ we had $n_k\simeq (Q/\lambda^2)k^{-d-2/3}$ (\ref{KZspin}). Let us now look at  strong turbulence, and argue for what we expect $n_k$ to be. Now the renormalized vertex $\Lambda$ is determined by $n_k$. Since for any $n_k$, the quantity $\mL_0\mL_2$ is positive for all or some directions of the wavevectors, we need to avoid the pole in \eqref{Lambda1234}. Both conditions  $\mL_0\mL_2=$const and    $\mL_0\mL_2=\mL_1^2$ give the same critical-balance scaling, 
 \be \label{CB00}
 n_k=Ak^{2-d},
 \ee
 up to logarithmic factors. 
  Let us check that this scaling corresponds to a constant flux solution. 
The critical balance scaling (\ref{CB00}) gives $\mL_0\simeq -A$ and $\mL_2\simeq -\lambda^2Ak_m^4$, where $k_m$ is the upper boundary of the critical-balance spectrum, whose value will be established later. We will assume that with this choice of $n_k$ the denominator in $\Lambda$ becomes vanishingly small at small $k$. On the basis of analyticity in $\o_k$, it is natural to assume that the denominator in \eqref{Lambda1234} goes to zero at $k\to0$ as $k^2/k_m^2$, i.e. one may imagine Taylor expanding $\mL_r$ to next order in $\o_1$, which is small relative to $\o_5$ (since the integrals in $\mL_r$ are UV divergent at leading order). This then causes the interaction vertex to grow as 
\be \label{Lam12}
\Lambda\simeq \mL_2k_m^2/k^2\simeq  \lambda^2Ak_m^6/k^2~,
\ee
where we used that the dominant term in the numerator of $\Lambda$ at low $k$ is $\mL_2$, as the $\mL_0$ and $\mL_1$ terms are multiplied by $\lam_{12}$, which vanishes at low $k$. Inserting (\ref{Lam12}) into (\ref{fluxQ})  gives $Q\simeq \lambda^4k_m^{12}A^5$.  
This flux is independent of the wavenumber,  demonstrating (\ref{CB00}) is a stationary solution for the low $k$ modes.
\begin{figure}[t] 
\centering 
\subfloat[]{
\includegraphics[width=2in]{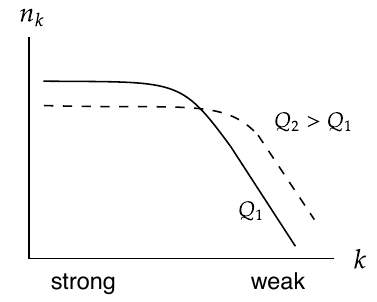}} \  \ \ \ \ \ \  \ \ \ \ \ 
\subfloat[]{
\includegraphics[width=1.9in]{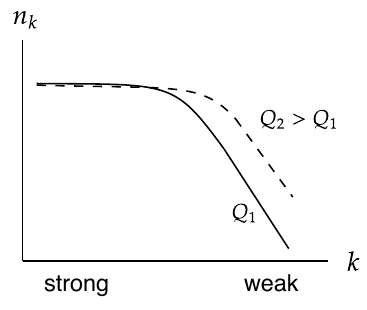}}
\caption{A sketch of $n_k$ for (a) spin waves (b) focusing nonlinear Schr\"odinger equation. In (a) $n_k$ is given by the weak turbulence scaling (\ref{KZspin}) at high $k$, while for low $k$ it is given by the critical balance scaling (\ref{CB0}), so increasing the flux increases the turbulence level at high $k$ but decreases it a low $k$. For the focusing nonlinear Schr\"odinger equation, studied in \cite{Rosenhaus:2025tjx, RF2}, the spectrum at low $k$ has the same scaling with $k$, but the prefactor is flux independent.  
} \label{Ploop}
\end{figure}

 Let us find the prefactor $A$. Inverting the expression for the flux gives $A \simeq (Q/\lam^4 k_m^{12})^{1/5}$. 
We insert this into the terms in the denominator of $\Lambda$,  and estimate $\mL_1^2\simeq \mL_0\mL_2\simeq A^2\lambda^2 k_m^4\simeq (Q\lambda/k_m^2)^{2/5}$. This dimensionless parameter should not be small, since if it were we would still be in the weak turbulence regime.
As a result, the cutoff momentum $k_m$ where critical balance stops being valid should be smaller than $\sqrt{Q \lam}$. This value is not new -- indeed, 
recall that $ \sqrt{Q\lambda}=k_*$ is the wavenumber where $\epsilon(k_*)\simeq 1$, see (\ref{39}). It is the wavenumber for which even the spectrally local interaction is strong. We thus conclude that the critical-balance state starts at $k_m=k_*=\sqrt{Q\lambda}$: for $k<k_*$ we have
\be 
n_k\simeq {k^{2-d}\over \lambda k_*^2}= {k^{2-d}\over \lambda^2Q}\ .\label{CB0}
\ee
Remarkably, the occupation numbers in the strong-turbulence region decrease with increasing  flux $Q$! This is evidently the result of spectral nonlocality and interaction enhancement: Increasing flux  increases both the transition wavenumber and the renormalized vertex, requiring lower occupation numbers. 
This decrease of $n_k$ with the growth of $Q$ saturates when  $k_*$ reaches $k_p$, i.e., if  strong turbulence starts already from the pumping scale. In this case, the spectrum is flux-independent, but explicitly dependent on the pumping scale: 
\be n_k\simeq {k^{2-d}\over \lambda k_p^2}\ .\label{CB1}\ee

Let us compare (\ref{CB0}) with the strong turbulence scaling for the focusing large $N$ nonlinear Schr\"odinger equation  \cite{Rosenhaus:2025tjx, RF2}. The vertex $\Lambda$ is given by (\ref{45}), and the argument for critical balance scaling proceeds similarly to what we just saw for spin waves --- one assumes an $n_k$ close to (\ref{CB00}), so that at low $k$, $\mL_-\simeq 1 + k^2$, and correspondingly
\be \label{LamNLS}
\Lambda \simeq 1/k^2~,
\ee
which gives constant flux. Unlike (\ref{Lam12}),  the $\Lambda$ here  has no $\mL_2$ dependence; in particular, it doesn't depend on the prefactor $A$ of $n_k$. The spectrum is thus flux independent, $n_k \simeq k^{2-d}$. The difference in behavior of $n_k$ for spin-waves and the nonlinear Schr\"odinger equation is illustrated in Fig.~\ref{Ploop}. 

Returning to spin wave turbulence, let us briefly comment on the behavior of $n_k$ at intermediate $k$, that lie in between the critical balance regime at low $k$ and the weak turbulence scaling at high $k$. In our discussion we assumed that $k_*<k_p$, so there exists an intermediate wavenumber $\mK \equiv k_* \(k_p/k_*\)^{2/3}$ where the one-loop correction term $\mL_2/\lambda_{12}$  is comparable to the leading-order term, see (\ref{WWC}). 
The more flat spectrum \eqref{CB0} starts at  $k_*$. The behavior of the spectrum  in the transition region $k_*{<}k{<}\mK$ must depend on the relative contributions to the the loop integrals of the regions of small and large wavenumbers. Steepening by the UV term $\mL_2/\lambda_{12}$ in the numerator is counteracted by the denominator term $\mL_0\mL_2$, for which the IR-divergent factor is $\mL_0\simeq 1/\lambda_{12}$, as long as  the IR cut-off is $k_*$. To summarize: for a long cascade, weak turbulence is realized for $k{>}k_* \(k_p/k_*\)^{2/3}$, the critical-balance state of strong turbulence is realized for $k<k_*$, and the scaling at intermediate $k_*{<}k{<}\mK$ requires a more detailed analysis which is left for future work.

\subsection{Strong turbulence for a class of models}

\subsubsection*{Models with sum-factorized interactions }
The discussion in the previous section can easily be generalized to a broader class of models, with interaction, 
\be
\lam_{1234} = - \lam \( (\v p_1 {\cdot} \v p_2)^{\beta/2} + (\v p_3 {\cdot} \v p_4)^{\beta/2}\)~.
\ee
Setting $\beta = 2$ gives spin waves, but we will consider general $\beta$ without affecting the result we will find below. In additional, we will take the dispersion relation to have the general scaling exponent $\al$,  $\omega_k=k^{\alpha}$. 
The weakly-turbulent (KZ) inverse cascade is then $n_k\simeq (Q/\lambda^2)^{1/3}k^{-d-2\beta/3+\alpha/3}$. For such $n_k$, the dimensionless coupling scales as $\epsilon_k\sim \lambda n_kk^{d-\alpha}\propto k^{(\beta-2\alpha)/3}$. Assuming that $2\alpha>\beta>0$, weak turbulence turns into a strong one below some $k_m$. We suggest that it is a critical-balance state. Up to logarithmic factors, the occupation numbers there are as follows:
\be 
n_k\simeq {k^{\alpha-d}\over \lambda k_m^\beta}\ .\label{generic}
\ee
For such a state, the loop integrals are UV divergent (cut off at $k_m$): $\mL_0\simeq -A$, $\mL_1 \sim -\lam A k_m^{\beta}$, $\mL_2\simeq -A \lam^2 k_m^{2\beta}$, where $A \simeq 1/\lambda k_m^\beta$. This gives $\mL_1^2 \simeq \mL_0 \mL_2 \simeq 1$, and correspondingly a vanishing denominator in $\Lambda_{1234}$ (\ref{Lambda1234}) as $k\to0$. 
The way it vanishes and $\Lambda_{1234}$ grows is determined by the frequencies in the denominators of the loop integrals. The next-order term gives corrections to $\mL_r $ proportional to $ \o_k/\o_{k_m} \sim k^{\al}/k_m^{\al}$. As a result, at low $k$ the renormalized interaction vertex behaves as,
\be 
\Lambda \simeq \mL_2 k_m^{\al}/k^{\al} \simeq \lam k_m^{\al+\beta}/k^{\al}\simeq \lambda k_m^\beta{\omega_{k_m}\over\omega_k}\ .\label{Lambda}
\ee
The interaction is enhanced inversely proportional to the kinetic energy of the quasiparticle. Inserting (\ref{generic},\ref{Lambda})  into the expression for the flux,
\be \label{fluxQQ}
Q\simeq \Lambda^2\frac{n_k^3k^{3d}}{\o_{k}} \simeq \frac{k_m^{2\al-\beta}}{\lam}
\ee
we see that the flux is indeed $k$-independent, so the strong-turbulence solution (\ref{generic}) is self-consistent. The critical balance is realized at $k<k_m=(\lambda Q)^{1/(2\alpha-\beta)}$, as follows from \eqref{fluxQQ}. That means that, like for spin waves, the spectral energy of strong turbulence {\it decreases} with the flux,
$n_k\propto k_m^{-\beta}\propto Q^{\beta/(\beta-2\alpha)}$. That decrease saturates when $k_m$ reaches $k_p$, so that strong turbuence starts already at the pumping scale.

Since the second powers of $\lambda$ were involved in the above consideration, the state \eqref{generic} is expected for both signs of $\lambda$.

\subsubsection*{Models with product-factorized interactions}
Consider now interactions of the form $\lam_{1234} = \lam(p_1 p_2 p_3 p_4)^{\beta/4}$, with $\o_k = k^{\al}$. The renormalized interaction vertex takes the form
\be \label{46v2}
\Lambda_{1234} = \frac{\lam_{1234}}{1- \mL_-}~, \ \ \ \ \mL_- =   2\lam\int\!\! d\vec{ p}_5 d\vec{ p}_6\,  (p_5 p_6)^{\beta/2} \frac{n_6{-}n_5}{\o_{16;35}{+}i\eps }\delta(\v p_{16;35})~,
\ee
which for $\beta = 0$ reduces to (\ref{45}) and for $\beta=2$ reduces to (\ref{46}). 

In the focusing case (negative $\lam)$, assuming a critical-balance solution, so that $\mL_-\sim 1+ {\rm O}(\o_k)$ and hence $\Lambda \propto k^{\beta}/\o_k$, we have that  for $n_k \simeq A k^{-\g}$, the flux is, 
\be
Q \simeq \Lambda^2\frac{n_k^3k^{3d}}{\o_{k}} \propto k^{2(\beta - \al)}  k^{3(d-\g)-\al}
\ee
For this to be $k$-independent we must have $n_k \propto k^{ \al -d- 2\beta/3}  $. Setting the UV-divergent loop integral $\mL_-$ (with cutoff $k_m$) to be equal to unity, we express prefactor $A$ via the cut-off $k_m$, so that the critical-balance state of an inverse cascade in the focusing case is as follows:
\be
n_k \simeq \frac{k^{ \al - \frac{2}{3} \beta -d}}{\lam k_m^{\beta/3}}
\ee
This is valid for $\al>1$ and $\beta>0$. The critical-balance solution for $\al<1$ is likely to be $n_k\propto k^{\alpha-d-\beta}$; each such case needs to be analyzed separately to establish the sign of the loop integral $\mL_-$, which converges both at IR and UV.

In the defocusing case (positive $\lam$), in the strong-turbulence regime we may drop the unity in the denominator of $\Lambda$, writing it as $\Lambda_{1234} \simeq \lambda_{1234}/\mL$. 
Self-consistently assuming that $\mL$ is IR divergent  for $n_k \simeq A k^{-\g}$ with IR cutoff $k_0$, we estimate $ \mL \simeq  A k_0^{d + \beta -\g} k^{-\a}$. Again we see that the renormalized repulsion is suppressed proportional to the frequency.
Flux constancy now requires $\gamma=d+(2\beta+\al)/3$, so that 
\be n_k\simeq Qk_0^{(\beta-\alpha)/3}k^{d+(2\beta+\al)/3}\,,\label{defocus}\ee
which is self-consistent with IR divergence when $\alpha>\beta>0$. 

Since the discussion in this subsection dealt only with the loop integrals diverging either at IR or UV, the conclusions are valid for arbitrary $\lambda_{1234}$ as long as it is factorized in the limit $p_1,p_3\ll p_2,p_4$. In some important physical cases, this factorization is asymmetric:  $\lam_{1234} = \lam(p_1 p_3)^{\beta_1/4}( p_2 p_4)^{\beta/4}$ with $\beta_1>\beta/2$ (notably, for gravity waves on water surface, $\beta_1=2, \beta=3$). In such cases, the loop integrals may converge at both IR and UV, which then requires the case-specific detailed analysis to establish effective interaction renormalization and resulting strong-turbulence spectrum.

\section{Discussion} \label{Sec5}

Our approach has been straightforward: kinetic equations for weakly interacting wave can be derived perturbatively in the nonlinearity. We analyzed  the next-to-leading order kinetic equation in the vicinity of the weak wave turbulent (Kolmogorov-Zakharov) state for inverse cascades in three classes of models.  For spin waves,  there is a next-to-leading order term (arising from one-loop vertex renormalization) that is dominated by large wavenumbers, and is cutoff by the pumping scale $k_p$. This introduces a dependence on the UV scale and breaks the universality of wave turbulence. 

In the original conception of quantum field theory, applied to particle physics, it was necessary for theories to be UV complete -- independent of the UV cutoff (such as a lattice regulator) which is a theoretical artifact, discarded at the end of the calculation. In the modern conception of effective field theories, Lagrangians are tuned so as to reproduce the observable long distance physics of interest, with little regard for the UV behavior. An inverse turbulent cascade is different: pumping is done in the UV where the cascade originates and where we have the most analytic control. One is not at liberty to tune the Hamiltonian, nor should one fear the UV cutoff dependence -- the pumping scale is not an artifact; without pumping there is no turbulent state. The belief that $k_p$ can be sent to infinity is an illusion, an artifact of the universality of Kolmogorov-Zakharov that arrises from only looking at very weak nonlinearity.

In Sec.~\ref{largeNsec} we studied the large $N$ generalization of just the quartic interaction sector of spin waves. If our goal is to truly study the behavior of spins at strong nonlinearity, we should generalize the Heisenberg ferromagnet Hamiltonian (\ref{A1}). Namely, we should take the same Hamiltonian, 
but with $\v S$ transforming in the fundamental representation of $SO(N)$, so that the Poisson brackets are given by 
$
\{S_a, S_b\} = f_{a bc} S_c
$,
where $f_{abc}$ are the structure functions of $SO(N)$. The IR behavior of this theory may not have the same IR behavior as (\ref{NHam}). The essential difference is that the spin variables $\v S$ are bounded from ever getting too large, whereas $\v \psi$ has no such constraint. We save the study of the kinetic theory of (\ref{A1}) for future work. 

Assuming that the critical balance solution (\ref{CB0}) holds for true spin waves,  the
 inability to create occupation numbers exceeding \eqref{CB0} is likely connected to wave breaking, as is the case in other contexts in which the critical balance solution appears:  self-focusing  in the Nonlinear Schr\"odinger Equation and, hypothetically, whitecaps for water waves \cite{RF2,FR}. Once 
 the dimensionless nonlinearity parameter, $\lambda n_kk^{d}$ -- which is the rms precession angle -- reaches an order-one value, we expect wave breaking, which presumably corresponds to the overturning of spins and the creation of localized nonlinear perturbations. The universal spectrum  \eqref{CB0} would then be dominated by bound states of spin waves, for which the nonlinearity and dispersion are balanced on all scales.

The UV divergence makes the universal spectrum \eqref{CB0} different from its dimensional estimate. Indeed, the so-called universal spectral balance generally requires the nonlinearity parameter, $\epsilon_k\simeq \lambda_{1234}n_kk^d/\omega_k$,  to be $k$-independent   \cite{Phil, Goldreich,Newell, NS, RF2}, which  in our case would give the  incorrect estimate $n_k\simeq k^{-d}$. The scaling $n_k\propto k^{2-d}$ is the same as in the focusing nonlinear Schr\"odinger model, and coincides with equipartition (up to a logarithmic factor) in two dimensions \cite{Rosenhaus:2025tjx}.

An inverse  cascade in an isothermal plasma is expected to behave similarly to spin wave cascades  and will be analyzed elsewhere. Such a detailed application of this new approach to  plasma turbulence could be important  for thermonuclear studies:  it may be necessary to go beyond  the weak turbulence approximation, even for weak nonlinearity,  despite  current belief to the contrary.

We conclude returning to one of the themes of the paper --- how nonlocality enhances nonlinearity --- which hinges on the divergences of loop integrals. In the product-factorized models (including the previously described nonlinear Schr\"odinger equation \cite{RF2,Rosenhaus:2025tjx}),   there is a single one-loop integral: it may converge, or it may diverge either in the UV or in the IR, but not both. For the spin wave interaction, there are effectively three separate integrals, $\mL_0, \mL_1, \mL_2$, which have different convergence properties. At one-loop level, the sum of these integrals appears (\ref{48}). In the large $N$ model, summing all the bubble diagrams produces a product $\mL_0 \mL_2$ of two of these integrals, see (\ref{Lambda1234}). For the KZ state one of the integrals ($\mL_0$) diverges at small wavenumbers, and the other one ($\mL_2$) diverges at large wavenumbers. That means that nonlocality coming from both ends of the cascade which is enhancing nonlinearity. While this is perhaps the  first such known case in turbulence, it  likely occurs in the wave turbulence of many other systems.

\sss*{Acknowledgments} 
We thank  G.~Eyink, D.~Schubring, and M.~Smolkin for helpful discussions. JL and VR are supported by NSF grant 2209116 and by BSF grant 2022113. GF is supported by the Excellence Center at WIS, the ISF grant  146845, and the BSF grant 2024019.

\appendix

\section{One-loop terms for spin waves} \label{AppOne}
In this appendix we evaluate  the one-loop terms in the kinetic equation (\ref{onelevelKE}) for the quartic spin wave interaction (\ref{lam1}), $\lam_{1234} = - \lam (\v p_1{\cdot} \v p_2 +\v p_3 {\cdot } \v p_4)$. Inserting this interaction into $\mL_{\pm}$ appearing in (\ref{onelevelKE}) gives,  
\bea
\mL_+ &=& -2\lam\!\int\!\! d \vec{p}_5 d \vec{p}_6 \frac{(\v p_1{\cdot} \v p_2{+}\v p_5{\cdot} \v p_6)^2}{\v p_1{\cdot }\v p_2} \frac{ n_5}{\o_{12;56}  }\delta(\v p_{12;56})~, \\ 
\mL_-&=& -\lam\!\int\!\! d\vec{ p}_5 d\vec{ p}_6  \frac{(\v p_3{\cdot} \v p_5{+}\v p_1{\cdot} \v p_6)(\v p_4{\cdot} \v p_6{+}\v p_2{\cdot} \v p_5)}{\v p_1{\cdot }\v p_2} \frac{n_6{-}n_5}{\o_{16;35} }\delta(\v p_{16;35})~, \label{A3}
\eea
where we used that  since $\mL_+$ is multiplied by momentum and frequency conserving delta conserving, we may replace
$\v p_3 {\cdot} \v p_4 \rightarrow \v p_1{\cdot} \v p_2$. 
Further, using that 
\be
2\v p_5{\cdot} \v p_6 = (\v p_{5}{+}\v p_6)^2 {-} p_5^2 {-} p_6^2= (\v p_{1}{+}\v p_2)^2 {-} p_5^2 {-} p_6^2 = \o_{12;56}{+}2 \v p_1{\cdot} \v p_2~, 
\ee
we get
\be \label{A4}
\mL_+  =\frac{-2\lam}{\v p_1{\cdot }\v p_2}\!\int\!\! d \vec{p}_5 d \vec{p}_6  \frac{(2 \v p_1{\cdot} \v p_2{+}\frac{1}{2}\o_{12;56})^2 }{\o_{12;56}  }n_5\delta(\v p_{12;56}){ =} {-}2\lam\!\int\!\! d \vec{p}_5 d \vec{p}_6  \(\frac{4 \v p_1{\cdot} \v p_2}{\o_{12;56}} {+}\frac{3}{2} {-}\frac{p_5^2}{2\v p_1{\cdot }\v p_2}\) n_5 \delta(\v p_{12;56})
\ee
where in the second equality we used $\frac{\o_{12;56}}{2} = \v p_5{\cdot}(\v p_1{+}\v p_2) - \v p_1{\cdot} \v p_2- p_5^2$ and the fact that the angular integral causes $ \v p_5{\cdot}(\v p_1{+}\v p_2) $ to integrate to zero. 
There are three terms in the integrand for $\mL_+$. The last term dominates in the UV, 
\be \label{C5}
\mL_+\Big|_{UV} = \frac{\lam}{\v p_1{\cdot }\v p_2}\mE~,
\ee
where $\mE$ is given by (\ref{mE2}) for the KZ state; it depends on the UV cutoff. On the other hand, the first two terms in the integrand in (\ref{A4}) dominate in the IR. In the limit of small $p_5$, $\o_{12;56}\rightarrow - 2 \v p_1{\cdot} \v p_2$, and we get, 
\be \label{C6}
\mL_+\Big|_{IR} = \lam \mN~,
\ee
where $\mN$ is given by (\ref{mN2}) for the KZ state; it depends on the IR cutoff. 
The first term in the integrand in (\ref{A4})  gives, in addition, a cutoff independent contribution. The angular integral in (\ref{A4}) is straightforward to evaluate, see \cite{RF2}. In particular, in three dimensions $\mL_+$ becomes, 
\be \label{C7}
\mL_+ =  - 8\pi\lam \int_{k_*}^{k_p}\!\! \frac{dq}{(2\pi)^3} q\, n_q \(\frac{  \v p_1{\cdot} \v p_2}{|\v p_1{+}\v p_2| } \log\Big|\frac{q^2 {-} q |\v p_1{+}\v p_2| {+}\v p_1{\cdot} \v p_2}{q^2 {+} q |\v p_1{+}\v p_2|{+}\v p_1{\cdot} \v p_2}\Big| + \frac{3}{2} q-\frac{q^3}{2\v p_1{\cdot} \v p_2}\)~,  \ \ \ d=3~.
\ee

Now let us look at $\mL_-$ in (\ref{A3}). The integral is convergent in the UV: taking large $p_5$ and Taylor expanding $n_5 - n_6$ and $\o_{16;35}$, as outlined in \cite{Rosenhaus:2025bgy}, and letting  $n_k \sim k^{-\g}$, we get, 
\be
\mL_-\Big|_{UV} = \frac{2 \g \lam}{\v p_1{\cdot} \v p_2}  \int^{k_p} \frac{d^d p_5}{(2\pi)^d} p_5^{-\g-2} (\v p_1{+}\v p_3){\cdot}\v p_5(\v p_2{+}\v p_4){\cdot}\v p_5\sim k_p^{d-\g}~,
\ee
which decays for $\g = d{+}2/3$. On the other hand, in the IR $\mL_-$ behaves as, 
\be \label{C9}
\mL_- \Big|_{IR} =  \lam\frac{(\v p_1{-}\v p_3)^2}{2\v p_1{\cdot} \v p_2} \mN~.
\ee
To get the cutoff independent contributions of $\mL_-$, one can evaluate the integrals in (\ref{A3}) if one wishes, however the angular integral is more involved than in the case of $\mL_+$, as there is dependence on the relative angle between two angles, such as between $\v p_5$ and both $\v p_2$ and $\v p_3$. 

Combining the cutoff dependent terms: (\ref{C5}), (\ref{C6}), and (\ref{C9}), and using that $\mL$ in (\ref{oneLQ}) is defined as $\mL \equiv 2 \mL_+ + 4 \mL_-$, gives  the equation (\ref{324}) quoted in the main body. 
\\[0pt]

In our study of the large $N$ kinetic equation, (\ref{Lambda1234}), we encountered $\mL_0, \mL_1, \mL_2$, which are pieces of the loop integral $\mL_+$. 
Specifically, in the KZ state the integral $\mL_0$ is dominated by the IR cutoff $k_0$, $\mL_2$ is dominated by the UV cutoff $k_p$, and $\mL_1\sim \eps_k$ is independent of $k_0$ and $k_p$. 
 Writing just the $k_0$ and $k_p$ dependence, $\mL_0$ and $\mL_2$ behave as 
\bea \label{mL0}
\!\!\!\!\!\!\!\!\!\!\!\! \mL_0 \!\! &=&\!\! 4\int_{k_0}d \v p_5 d \v p_6 \frac{n_5}{\o_{34;56}{+}i\eps} \delta(\v p_{12;56}) = -\frac{2}{\v p_1{\cdot} \v p_2} \mN= - 3 S_d \(\frac{Q}{\lam^2}\)^{1/3} \frac{k_0^{-2/3}}{\v p_1{\cdot} \v p_2}\sim -\frac{1}{\lam} \(\frac{k_*}{k_0}\)^{2/3}\!\!  \frac{1}{\v p_1{\cdot } \v p_2} \ \ \ \ \ \ \ \  \ \ \\ \label{mL2}
\mL_2 &=&\!\!\!\! 4\lam^2\int^{k_p} d \v p_5 d \v p_6\, (\v p_5{\cdot} \v p_6)^2 \frac{n_5}{\o_{34;56}{+}i\eps} \delta(\v p_{12;56})  =  -2\lam^2 \mE=  -\frac{3}{2}S_d \(Q \lam^4\)^{1/3} k_p^{4/3}\sim - \lam k_*^2 \(\frac{k_p}{k_*}\)^{4/3}~ \ \ \ \ \  \ \ \ \ 
\eea
The product $\mL_0 \mL_2$ that appears in the  denominator of (\ref{Lambda1234}) behaves as, 
\be \label{415}
\mL_0 \mL_2 \sim   \frac{(\lam Q)^{2/3}}{\v p_1{\cdot} \v p_2} \(\frac{k_p^2}{k_0}\)^{2/3}\sim \frac{k_*^2}{\v p_1{\cdot} \v p_2} \(\frac{ k_p^2}{k_0 k_*}\)^{2/3}\sim \frac{\mK^2}{\v p_1{\cdot} \v p_2} \(\frac{k_*}{k_0}\)^{2/3}~.
\ee

\bibliographystyle{utphys}

\end{document}